\documentclass[pra,preprint]{revtex4}
\usepackage{subeqnarray,amsfonts,amssymb,amsbsy,bm,graphicx}

\def\que#1#2{\displaystyle\frac{#1}{#2}}
\def\pri{^\prime}

\let\De=\Delta
\let\al=\alpha

\let\ra=\rightarrow
\let\va=\varphi

\let\la=\lambda

\begin{document}
\let\disp=\displaystyle
\let\ti=\times
\let\si=\sigma
\let\Om=\Omega
\let\om=\omega
\let\te=\theta
\let\ti=\times
\let\ov=\overrightarrow
\let\ove=\overrightarrow

\title{Interactions
between zero-point radiation and electrons}

\author{Rafael Alvargonz\'alez}
\affiliation{Universidad Complutense, 28040 Madrid, Spain}

\date{\today}

\begin{abstract}
    Knowing the magnitude of the energy flow
    inherent to zero-point radiation allows us to approach the
    question of its possible interaction with particles of matter. Its
    photons are not different from the rest, and must in principle be
    subject to the Compton effect and the Klein-Nishima-Tann formula
    for its cross section. On this assumption, it is shown here that
    zero-point radiation may be powerful enough to explain
    Poincar\'e's tensions and to supply an efficient cause for
    gravitation. This could be only the case if the classic radius of
    the electron measures $8.143375\times10^{20}q_\la$, where $q_\la$
    is the minimum wavelength for electromagnetic radiation, and if
    the wavelength of the most energetic photon in the actual
    zero-point radiation is $5.275601\times10^{27}q_\la$. To the first
    of these numbers there corresponds the energy $3.5829
    \times10^{23}$~MeV for the photon whose wavelength is $1q_\la$.
    This gives also the relation $q_\la=(2 \pi \al)^{1/2}L_P$, where $L_P$
    is the Planck Length. Finally the relation between the force of
    gravity and the electrostatic force is explained by the equations
    obtained in this paper.
\end{abstract}

\maketitle

\section{Preliminary considerations on zero-point radiation}

Sparnaay's 1958 experiments [1] exposed the existence of zero-point
radiation, which Nernst had considered as a possibility in 1916. However,
Sparnaay was not looking for it, since he was only trying to check Casimir's
hypo\-the\-sis about the mutual attraction of two uncharged conductor plates
placed very close together [2]. This attraction should have disappeared when
the tempe\-rature approached absolute zero, but Sparnaay found that at
near that tempera\-ture there was still some attraction not accounted for by Casimir's
hypothesis, being independent of temperature and obeying a very simple law;
it is directly proportional to the surface of the plates, and inversely
proportional to the 4$^{\rm th}$ power of the distance $d$ between them.
Sparnaay observed this force when the plates were placed in a very complete
vacuum at near zero-absolute temperature. For a distance of $5\ti10^{-5}$ cm
between the plates, he was able to measure a force of
$0.196\ {\rm g cm s}^{-2}$, and deduced the formula
$$f=\que{k_s}{d^4};\qquad\hbox{where}\ \ k_s=1.3\cdot10^{-18}\
{\rm erg} {\rm cm}$$

In a near-perfect vacuum and at a temperature very near absolute zero, which
implies the absence of any ``photon gas", the phenomenon observed by Sparnaay
could only be produced by a radiation inherent to space. This could be only
the case if its spectrum is relativistically invariant, which could only
happen if its spectral distribution is inversely proportional to the cubes of
the wavelengths --in other words, if the number of photons of wavelength $\la$
which strike a given area within a given time is inversely proportional to
$(\la)^3$.

A function of spectral distribution which is inversely proportional to the
cubes of the wavelengths, implies a distribution of energies which is
inversely proportional to the 4$^{\rm th}$ power of the wavelengths, because
the energies of the photons are inversely proportional to the wavelengths. In
1969, Timothy H. Boyer [4] showed that the spectral density function of
zero-point radiation is
$$f_\va(\la)=\que1{2\pi^2}\que1{(\la_*)^3},\eqno{(1)}$$
where $\la_*$ is the number giving the measurement of wavelength $\la$.

This function produces the next, for the corres\-pon\-ding energies
$$E_\va(\la)=\que1{2\pi^2}\que{hc}\la\que1{(\la_*)^3}\eqno{(2)}$$
for $\la\ra0$, $E_\va(\la)\ra\infty$. There must therefore be a threshold for
$\la$, which will hereafter be designated by the symbol $q_\la$.

In the Sparnaay effect, the presence of a force which is inversely
proportional to the 4$^{\rm th}$ power of the distance $d$ between the
plates, while the distribution of energies is also inversely proportional to
the 4$^{\rm th}$ power of the wavelengths, leads us to infer that the cause of
the apparent attraction must lie in some factor which varies inversely with
the wavelength, and which causes a behaviour different from the photons of
wavelengths equal to or greater than $d$, since:
$$\sum^\infty_{n=d}Kn^{-5}\ra\que K{4d^4}$$

The factor in question turns out to be the proportion of photons which are
reflected, not absorbed; that is the coefficient of reflection $\rho$, while
the different behaviour is caused by the obstacle which the presence of each
one of the two plates offers to the reflection from the inner face of the
other one, of photons of wavelength equal or greater than ``$d$". As we know,
the energy which is transferred by a reflected photon is double that
transferred by one which is absorbed.

To simplify the following arguments, it is convenient to use the $(e,m_e,c)$
system of measurements in which the basic magnitudes are the quantum of
electric charge, the mass of the electron and the speed of light. In this
system, the units of length and time are respectively $l_e=e^2/m_ec^2$ and
$t_e=e^2/m_ec^3$.

The results of Sparnaay's experiments provide a link with reality which may be
enough to show the intensity of the energy flow belonging to zero-point
radiation. A recent paper [5] has shown that this flow is one which
corresponds to the incidence in an area $(q_\la)^2$ of one photon of
wavelength $q_\la$ every $q_\tau$, plus one photon of wavelength $2q_\la$
every $2^3q_\tau$, etc. up to one photon of wavelength $nq_\la$ every
$n^3q_\tau$, where $q_\tau=q_\la/c$ [6].
This radiation implies the energy flow
per $(q_\la)^2$ and $q_\tau$ which is given by
$$W_0=\que{hc}{q_\la q_\tau}\left\{1+\que1{2^4}+\cdots+\que1{n^4}\right\}=
\que{\pi^5}{45\al}(k_\la)^2\que{m_ec^2}{t_e},\eqno{(3)}$$
where $k_\la=l_e/q_\la=t_e/q_\tau$ [7].

This energy flow produces a force per $(l_e)^2$ which is given by
$$F_{(e,m_e,c)}=\que{\pi^5}{45\al}(k_\la)^4\que{m_ec}{t_e}$$
which when expressed in the c.g.s. system and per cm$^2$ is given by
$$F_{\rm (c.g.s.)}=3.409628\ti10^{34}(k_\la)^4\ \que{{\rm g cm}}{{\rm
s}^2}.$$

In cosmic rays, photons have been observed with energies of the order of
more than $10^{19}$~eV, [8] which implies that the value of $k_\la$ must be
greater than
$2.273\ti10^{10}$.
The value of $F_{\rm (c.g.s.)}$ is
therefore inmense. In consequence the
analysis of the possible interactions of zero-point radiation with the
elementary particles assumes enormous interest.

\section{Zero-point radiation and Poincar\'e tensions}

\begin{quote}
``If we have a charged sphere, all the electrical forces repel, and the
electron will tend to fly apart... The charge must be mantained over the
sphere by something which stops it from flying off. Poincar\'e was the first
to point out that this `something'\ must be allowed for in the calculation of
energy and momentum".

\hfill{{ The Feynman Lectures in Physics}, Vol. II, 28-1 to 28-14.}
\end{quote}

If the electron's mass and charge were distributed according to some spa\-tial
configuration, the repulsion of the charge against itself would tend to cause
the particle to fly apart. Seeing this, Poincar\'e suggested that something
must exist which can counteract this repulsion and this `something'\ has
therefore been named `Poincar\'e's tensions'. The possible configuration must
be one which has no favoured direction, which implies that it must have
spherical symmetry. Electrostatic repulsion would appear in the form of a
centrifugal force. To meet this, Poincar\'e's tensions would have to be
arranged as centripetal forces able to counteract it. When submitted to these
fields of opposing forces, the charge of the electron would tend to distribute
itself equally over a spherical surface. Zero-point radiation, which arrives
equally from all directions of space, provides centripetal energy flows
towards every imaginable spherical surface, and from these flows there could
be derived equally centripetal flows, to fill the role of Poincar\'e's
tensions. Sparnaay's experiments have allowed us to measure the intensity of
the energy flows inherent to zero-point radiation; Compton's experiments
discovered the laws which govern the phenomena of dispersal and energy
transfer produced by the interaction of photons with free particles and
matter. Finally the differential cross section for these phenomena follows the
Klein-Nishima-Tann formula. With this knowledge we can attempt to calculate
the centripetal force which could be produced by the zero-point radiation, on
an electron with its charge distributed equally over a spherical surface of
ra\-dius~$r_x$.

The collision of a photon with a free particle of matter, shown in Fig. 1,
produces the Compton effect. The colliding photon of energy $E_\la$ and
wavelength $\la$, loses part of its energy to the particle, and is diverted at
an angle $\te$ to its previous trajectory. When losing energy, it increases
its wavelength by an amount which is given by Compton's equation
$$\De\la=\la_c(1-\cos\te),$$
where $\la_c=h/m\,c$, where $m$ is
the said particle mass. For the electron, Compton's wavelength
$\De\la=\la_{ce}=h/m_ec=\que{2\pi}\al l_e$.

In the case of zero-point radiation, the free particle of matter, which is
subjected to equal forces from all directions, does not move, but suffers
compression towards its centre.

\begin{figure}
\centering
\resizebox{0.75\columnwidth}{!}{\includegraphics{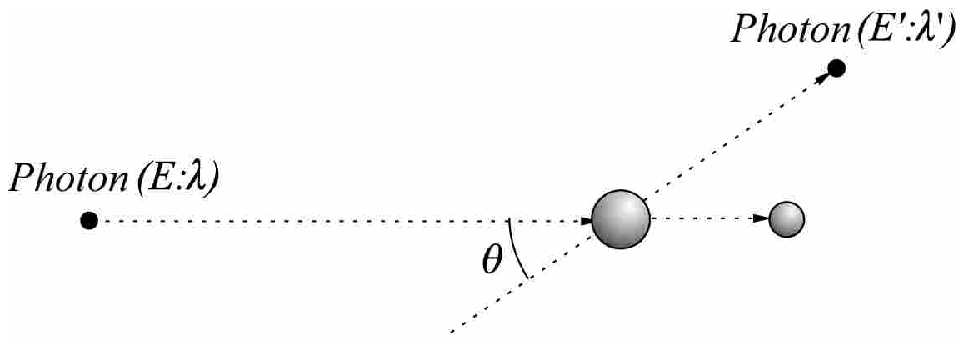}}
\caption{}
\end{figure}

The encounter of a photon with an electron does not always produce
the Compton effect; the differential cross section for this $d\si$
is given by the Klein-Nishima-Tann formula for dispersal with
electrons, which is
$$\que{d\si}{d\Om}=\que12(r_e)^2\left(\que{\om\pri}\om\right)^2
\left(\que{\om\pri}\om+\que\om{\om\pri}-\sin^2\te\right),$$
where $\te$ is the angle of dispersal, $\om$ the frequency of the arriving
photon, $\om\pri$ that of the dispersed photon, $r_e=l_e=e^2m_e^{-1}c^{-2}$
the classical radius of the electron, $d\si$ is the differential cross section
for the Compton effect, and $d\Om=2\pi\sin\te d\te$. By bringing in this
relation and remembering that $\que{\om\pri}\om=\que\la{\la\pri}$, we can
write
$$d\si=\pi(r_e)^2\left(\que\la{\la\pri}\right)^2
\left[\left(\que{\la\pri}\la+\que\la{\la\pri}\right)\sin\te-\sin^3\te\right]
d\te\eqno{(4)}$$

Taking all the foregoing into consideration, we also know that:
\begin{itemize}
\item The number of photons of zero-point radiation with wavelength $nq_\la$,
which converge from all directions in an area $(q_\la)^2$ is $1/n^3$ every
$q_\tau$.
\item The energy transferred to the electron by a photon of wavelength
$nq_\la=\la$, which after interacting is diverted at an angle $\te$ from its
trajectory, is

\end{itemize}
$$E=hc\left[\que1\la-\que1{\la\pri}\right]=
hc\left[\que1{nq_\la}-\que1{nq_\la+\que{2\pi}\al l_e(1-\cos\te)}\right].$$

Therefore, the energy transferred every $q_\tau$ by these photons of point-zero
radiation converging on an area $(q_\la)^2$ is given by
$$E_{T_n}=\que{\pi(r_e)^2}{n^3}hc\int^{\pi/2}_0
\left(\que1\la-\que1{\la\pri}\right)\left(\que\la{\la\pri}\right)^2
\left(\que\la{\la\pri}+\que{\la\pri}\la-\sin^2\te\right)
\sin\te\,d\te,\eqno{(5)}$$
where \
$$\la=nq_\la$$
$$\la\pri=nq_\la[1+2\pi k_\la(1-\cos\te)/\al n],$$
$$k_\la=l_e/q_\la,$$
whence we derive the following definite integrals:
$$\int^{\pi/2}_0\que{\la^2}{\la^{\prime3}}\sin\te d\te=
\que1{nq_\la}\que12\left[\que1{1+A}+\que1{(1+A)^2}\right],$$
$$\que1{nq_\la}\int^{\pi/2}_0\que{\sin\te d\te}{\la\pri}=
\que1{nq_\la}\que1A\ln(1+A),$$
$$-\que1{nq_\la}\int^{\pi/2}_0\que\la{\la^{\prime2}}\sin^2\te d\te=
\que1{nq_\la}\que1A\left[\que1{1+A}+\que2A-\que2{A^2}(1+A)\ln(1+A)\right],$$
$$-\que1{nq_\la}\int^{\pi/2}_0\que{\la^3}{(\la\pri)^4}\sin\te d\te=
-\que1{3nq_\la}\left[\que1{1+A}+\que1{(1+A)^2}+\que1{(1+A)^3}\right]$$
$$-\que1{nq_\la}\int^{\pi/2}_0\que\la{(\la\pri)^2}\sin\te d\te=
\que1{nq_\la}\que{-1}{nq_\la}\left(\que1{1+A}\right)\left(\que1{1+A}\right),$$
$$+\int^{\pi/2}_0\que{\la^2\sin^3\te d\te}{(\la\pri)^3}=\que1{nq_\la}\que1A
\left[\que1A-\que1{2(1+A)^2}-\que1{A^2}\ln(1+A)\right],$$
where $A=\que{2\pi}\al\que{k_\la}n$. The addition of this integrals gives
$$\begin{array}{c}
\disp\que1{nq_\la}\left[\que{l(1+A)}A-\que5{6(1+A)}+\que1{6(1+A)^2}-
\que1{3(1+A)^3}+\que3{A^2}+\right.\\ \\[-5pt]
\left.\disp+\que1{A(1+A)}-\que1{2A(1+A)^2}-
\que{(3+2A)}{A^3}l(1+A)\right]\end{array}\eqno{(6)}$$

Therefore we have:
$E_{T_n}=\pi(r_e)^2\que{hc}{nq_\la}[A]$; where
$$[A]=\left[\que{l(1+A)}A-\que5{6(1+A)}+\que1{6(1+A)^2}+\cdots\right]$$
as in (6).
Therefore $\que{E_{T_n}}{E_n}=\pi(r_e)^2[A]$; obviously $E_n=\que{hc}{nq_\la}$
and $\que{E_{T_n}}{E_n}<1$.

Since in the $(e,m_e,c)$ system $r_e=1$, this factor can be ignored and we can
write
$$\que{E_{T_n}}{E_n}=\pi[A],\eqno{(7)}$$
and
$$E_{T_n}=\que{hc}{nq_\la}\pi[A].\eqno{(8)}$$

A first attempt to solve the problem showed us that the energy flow (3)
would be excessive. Sparnaay's measurements allow us to deduce that, in the
case of photons with wavelength equal to or greater than $5\times10^{-10}$~cm,
the energy flow of the zero-point radiation agrees with the energy flow
corres\-pon\-ding to the incidence, in an area $(q_\la)^2$, of the photon of
wavelength $q_\la$ every $q_\tau$, plus another of wavelength $2q_\la$ every
$2^3q_\tau$, up to a photon of wavelength $nq_\la$ every $n^3q_\tau$. However
this does not mean that, at present, the photon with the shortest wavelength
in the zero-point radiation, has to be the photon of wavelength $1q_\la$. If we
suppose that its wavelength is $xq_\la$, the flow of energy to be taken into
account is:
$$W_x=\que{hc}{q_\la q_\tau}\sum^\infty_x\que1{n^4}=
\que{2\pi}{3\al}\que{(k_\la)^2}{x^3}\que{m_ec^2}{t_e},\ \ {\rm per}\ (q_\la)^2
\eqno{(9)}$$
instead of
$$W_0=\que{\pi^5}{45\al}(k_\la)^2\que{m_ec^2}{t_e},$$
as given by (3).

The condition $\que{E_{T_n}}{E_n}<1$ implies
$$\pi\left[\que{l(1+A)}A-\que5{6(1+A)}+\que1{6(1+A)^2}-\que1{3(1+A)^3}+
\que3{A^2}+\right.$$
$$\left.+\que1{A(1+A)}-\que1{2A(1+A)^2}-\que{2l(1+A)}{A^2}-
\que{3l(1+A)}{A^3}\right]<1$$

It is reasonable to suppose that the wavelength $xq_\la$ of the most ener\-ge\-tic
photon in the actual zero-point radiation is much greater than the wavelength
of the photon whose energy is $m_ec^2$, which is $\que{2\pi}\al l_e=
\que{2\pi}\al k_\la q_\la$. If this assumption does not cause any future
contradiction and leads to a value of $k_\la>2.273\times10^{10}$, which means
that the energy of the photon with wavelength $1q_\la$ is greater than
$10^{19}$~eV, it must be considered as reasonable. Accordingly
$A=\que{2\pi}\al\que{k_\la}x<1$, and we can write:
$$l(1+A)=A-\que12A^2+\que13A^3+\cdots+(-1)^{m+1}\que{A^m}m$$

Besides, we have for any value of $A$:
$$\begin{array}{l}
(1+A)^{-1}=1-A+A^2-A^3+\cdots+(-1)^mA^m,\\[+4pt]
(1+A)^{-2}=1-2A+3A^2-4A^3+\cdots+(-1)^m(m+1)A^m,\\[+4pt]
(1+A)^{-3}=1-3A+6A^2-10A^3+\cdots+(-1)^m\que{(m+1)(m+2)}2A^m.
\end{array}$$

By introducing these series into $[A]$, as in (6), we obtain:
$$\begin{array}{c}
\disp [A]=\que7{12}A-\que{11}{10}A^2+\que{37}{20}A^3+\cdots+\\ \\[-4pt]
\disp +(-1)^m
\left[\que1{m+1}+\que2{m+2}-\que3{m+3}-1-\que{m(m-1)}6\right]A^m
\end{array}\eqno{(10)}$$

To obtain the total energy transferred each $q_\tau=t_e/k_\la$ to an area
$(q_\la)^2$, we can write:
$$E_{T_x}=\sum^\infty_x\que{E_{T_n}}{n^3}.$$
From (8) we have
$$E_{T_n}=\que{hc}{nq_\la}\pi\ [A]=\que{2\pi^2}\al\que{k_\la}n[A]m_ec^2$$
whence
$$E_{T_x}=\que{2\pi^2}{\al}k_\la\left(\sum^\infty_x\que{[A]}{n^4}\right)m_ec^2$$
each $q_\tau$.

Therefore the energy flow per $(q_\la)^2$ each $t_e$ is:
$$\begin{array}{l}
\disp W_{T_x}=\que{m_ec^2}{t_e}\que{2\pi^2}\al(k_\la)^2
\sum^\infty_x\que1{n^4}\left\{\que7{12}A-\que{11}{10}A^2+\cdots+\right.\\ \\
\disp\left.+(-1)^m\left[
\que1{m+1}+\que2{m+2}-\que3{m+3}-1-\que{m(m-1)}6\right] A^m\right\}
\end{array}\eqno{(11)}$$
to an area $(q_\la)^2$. Since $A=\que{2\pi}\al\que{k_\la}n$ we can write:
$$\!\!\!\begin{array}{l}
\disp W_{T_x}=\que{m_ec^2}{t_e}\que{2\pi^2}\al(k_\la)^2
\left[\que7{12}\que{2\pi k_\la}\al\sum^\infty_x\que1{n^5}-\que{11}{10}
\left(\que{2\pi k_\la}\al\right)^2\sum^\infty_x\que1{n^6}+\cdots+\right.\\ \\
\disp\left.+(-1)^{m-1}
\left\{\que1{m+1}+\que2{m+2}-\que3{m+3}-1-\que{m(m-1)}6\right\}
\left(\que{2\pi k_\la}\al\right)^m\sum^\infty_x\que1{n^{m+4}}\right]
\end{array}\!\!\!\eqno{(12)}$$
to an area $(q_\la)^2$, whence
$$W_{T_x}=\que{m_ec^2}{t_e}\que{2\pi^2}\al(k_\la)^2
\left[\que7{48}\,\que{2\pi k_\la}\al\que1{x^4}-\que{11}{50}
\left(\que{2\pi k_\la}\al\right)^2\que1{x^5}\cdots+\right.$$
$$\left.+T_m\left(\que{2\pi
k_\la}\al\right)^m\que1{x^{m+3}}\right]$$
to an area $(q_\la)^2$; where
$$T_m=(-1)^{m-1}\left[\que1{m+1}+\que2{m+2}-\que3{m+3}-1-
\que{m(m-1)}6\right]\que1{m+3}$$

By introducing $B=\que{2\pi}\al\left(\que{k_\la}x\right)$ we can write:
$$\!\!\!W_{T_x}=\que{m_ec^2}{t_e}\que{2\pi^2}\al\que{(k_\la)^2}{x^3}
\left(\que7{48}B-\que{11}{50}B^2+\cdots+T_mB^m\right)\eqno{(13)}$$
to an area\ $(q_\la)^2$, in such a way that
$$\que{W_{T_x}}{W_x}=3\pi[B]_m,\eqno{(14)}$$
where we used
$$\{B\}_m=\left(\que7{48}B-\que{11}{50}B^2+\cdots+T_mB^m\right).$$

Fig. 2 shows an electron configured as a spherical surface as a result of the
equilibrium between centrifugal forces of electrostatical repulsion and
centripetal forces derived from the interaction with zero-point radiation. If
this interaction happens with zero-point radiation falling on an area
$(q_\la)^2$ situated on $N$ on the surface of the electron, and coming from all
directions of the half space defined by the tangent plan at $N$ and opposite
to $O$, the total energy flow over the said area will be $W_x/2$, and the
total energy flow transferred to the electron will be $\que{W_{T_x}}2$.

\begin{figure}[h]
\centering
\resizebox{0.75\columnwidth}{!}{\includegraphics{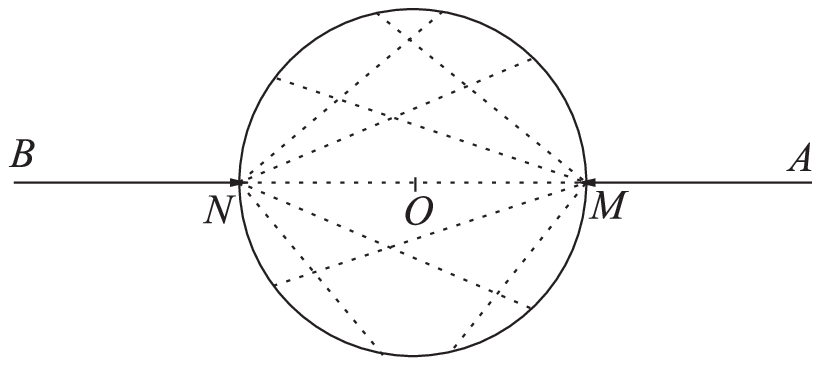}}
\caption{}
\end{figure}

This energy flow produces a force $F=\que{W_{T_x}}{2c}$. If we imagine a half
sphere whose center is at $N$ and whose radius has a measure equal to $F$ the
components according the direction $\ov{NO}$ of the forces $F$
coming fron the referred half space have the same measurements as the
distances of the points of the surface of the said sphere to the tangent plant
at $N$ and the number which expresses their sum is that which measures the
volume of the said half sphere; i.e. $\que{2\pi}3F^3$. Now, in the present
case we must divide this number by the area of the said half sphere, because
the total force $\ov{F_{NO}}$ is the sum of those which come
from all directions. To each point of this area there corresponds one of those
directions. Therefore the sum of the components according the radius at the
point of arrival of the forces which strike in an area $(q_\la)^2$ is given by
$$F_{NO}=\que{(2/3)\pi F^3}{2\pi F^2}=\que F3=\que{W_{T_x}}{6c},$$
which gives
$$F_{NO}=\que{\pi^2}{3\al}\que{(k_\la)^2}{x^3}[B]_m\que{m_ec}{t_e}\eqno{(15)}$$

The electrostatic repulsion for the fraction $e/4\pi(r_x)^2(k_\la)^2$ of the
charge $e$ which corresponds to a fraction $(q_\la)^2$ of a sphere with radius
$r_xk_\la(q_\la)$ is
$$F_e=\que e{4\pi(r_x)^2(k_\la)^2}e\que1{(r_x)^2(l_e)^2}=
\que1{4\pi(r_x)^4(k_\la)^2}\que{m_ec}{t_e}\eqno{(16)}$$

By equalising (15) and (16) we obtain
$$x^3=\que{4\pi^3}{3\al}(k_\la)^4(r_x)^4[B]_m\eqno{(17)}$$

\section{Behaviour of two electrons immersed in zero-point radiation}

\begin{figure}[b]
\centering
\resizebox{0.75\columnwidth}{!}{\includegraphics{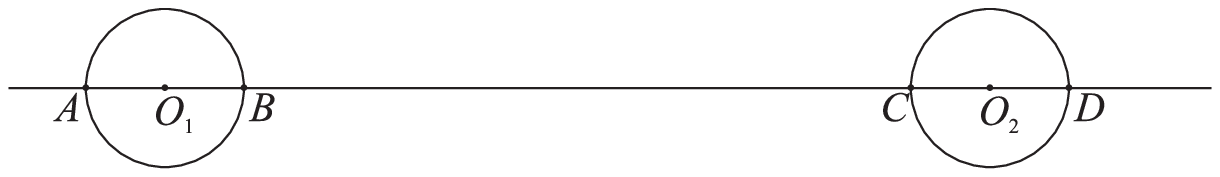}}
\caption{}
\end{figure}

Figure 3 shows two electrons whose centers are $d  l_e$ apart. A fraction of
the zero-point radiation flow which reaches the electron whose center is at
$O_1$ is spent to equalize the Poincar\'e's tensions at its surface. According
to (15) this fraction implies the force.
$$F_P=4\pi(r_x)^2 \que{\pi^2}{3\al}\que{(k_\la)^2}{(x)^3}[B]_m
\que{m_ec}{t_e}.$$
Therefore, the zero-point flows which emerge after interacting with this
particle imply forces towards $O_1$ whose intensities at the distance $d  le$
from this point would be equal to $F_P/3\pi d^2$. The same occurs with the
zero-point ra\-dia\-tion which interacts with the electron at $O_1$ after having
interacted with the electron at $O_2$. Both reductions produce an apparent
attraction given by
$$F_d=\que{2\pi^2(r_x)^2}{3\al d^2}\que{(k_\la)^2}{(x)^3}[B]_m
\que{m_ec}{t_e}\eqno{(18)}$$

The expression of $G$ in the $(e,m_e,c)$ system is $G=G_e\left(\que
e{m_e}\right)^2$, where $G_e=2.399998\times10^{-43}$.

The attraction between two electrons which are $d  l_e$ distant is:
$$F_G=\que{G_e}{d^2}\que{m_el_e}{t_e}.\eqno{(19)}$$

By equalising (18) and (19) we obtain
$$x^3=\que{2\pi^2}{3\al}\que{(k_\la)^2(r_x)^2[B]_m}{G_e}\eqno{(20)}$$

From (17) and (20) we obtain
$$k_\la r_x=\left[\que1{2\pi G_e}\right]^{1/2},$$
where $k_\la r_x$ is the length of the radius of the electron and
$k_\la=\que{l_e}{q_\la}$ the number of $q_\la$ by one $l_e$. The quantum
character of $\left(\que{e^2}{c^2}\right)^2=m_el_e$ requires that if $m_e=1$,
$r_x=Nl_e$, where $N$ is a digit.

For $N=1$ we have
$$k_\la=\left(\que1{2\pi G_e}\right)^{1/2}.\eqno{(21)}$$

The Planck length $L_P=\left(\que{hG}{c^3}\right)^{1/2}$, expressed in
the $(e,m_e,c)$ system is
$L_P=\left(\que{G_e}\al\right)^{1/2}l_e$. Therefore
$q_\la=\que{l_e}{k_\la}=( 2 \pi \al)^{1/2}L_P = 0.214276 L_P$.

$L_P$ is bigger than $q_\la$, on the same way that
$h=\que{2\pi}\al\que{e^2}c$ is bigger than $\que{e^2}c$.
Notwithstanding  the real wavelength of the photon of maximum energy is
$q_\la$ and not $L_P$. The quantum of the magnitude $\que{ML^2}T$ is bigger
than the easy combination $\que{e^2}c$. The quantum $q_\la$ can be expressed
as $q_\la=\left(\que{2 \pi G   e^2}{c^3}\right)^{1/2}$, which is minor
than the easy combination $\left(\que{G e^3}{\alpha c^3}\right)^{1/2}$.

\section{Conclusions}

The minimum value of $r_x$ is 1, and for it:
$q_\la=(2 \pi \al)^{1/2}L_P$ and
$$k_\la=\left(\que1{2\pi G_e}\right)^{1/2}=8.143375\times10^{20}$$
$$x=5.257601\times10^{27}$$
$$z=\que{k_\la}x=1.548877\times10^{-7}$$

The wavelengths of the most energetic photons observed in cosmic rays are more
than $10^5q_\la$.

The relation between the force of gravitation on the electrostatic
for\-ce~is:
$$\que G{e^2}=\que1{2\pi(k_\la)^2}=2.4000001 \times10^{-43}$$

\section*{References}

[1] M. J. Sparnaay, Physica {\bf 24}, 51 (1958).

[2] H. B. J. Casimir, J. Chimie Phys. {\bf 46},  497 (1949).

[3] P. W. Milonni, \textit{The Quantum Vacuum} (Acaddemic Press, New York, 1994).

[4] T. H. Boyer, Phys. Rev. {\bf 182}, 1374 (1969).

[5] R. Alvargonz\'alez, LANL arXiv: physics/031027

[6] Note that Sparnaay's experiments do not imply that
$q_\la$ must be the wavelength of the most energetic
photon in zero-point radiation.

[7] This is the flow by $(q_\la)^2$ and $t_e$.

[8] P. Blas, LANL arXiv: astro-ph/0206505

[9] I. Solokov: \textit{Quantum Electrodynamics} (Mir, Moscow, 1981)

\end{document}